# A Demonstration of Accurate Wide-field V-band Photometry Using a Consumer-grade DSLR Camera

**Brian K. Kloppenborg**
*Department of Physics and Astronomy, University of Denver, 2112 East Wesley Avenue, Denver, CO 80208; bkloppen@du.edu*

**Roger Pieri**
*37 C rue Charles Dumont, 21000, Dijon, France; roger.pieri@wanadoo.fr*

**Heinz-Bernd Eggenstein**
*Vogelbeerweg 34, 31515 Wunstorf, Germany; heinz-bernd.eggenstein@vspb.de*

**Grigoris Maravelias**
*Physics Department, University of Crete, GR-71003 Heraklion, Crete, Greece; gmaravel@physics.uoc.gr*

**Tom Pearson**
*1525 Beachview Drive, Virginia Beach, VA 23464; tandjpearson@verizon.net*

**Abstract**   The authors examined the suitability of using a Digital Single Lens Reflex (DSLR) camera for stellar photometry and, in particular, investigated wide field exposures made with minimal equipment for analysis of bright variable stars. A magnitude-limited sample of stars was evaluated exhibiting a wide range of ($B–V$) colors taken from four fields between Cygnus and Draco. Experiments comparing green channel DSLR photometry with VT photometry of the Tycho 2 catalogue showed very good agreement. Encouraged by the results of these comparisons, a method for performing color-based transformations to the more widely used Johnson V filter band was developed and tested. This method is similar to that recommended for Tycho 2 VT data. The experimental evaluation of the proposed method led to recommendations concerning the feasibility of high precision DSLR photometry for certain types of variable star projects. Most importantly, we have demonstrated that DSLR cameras can be used as accurate, wide field photometers with only a minimal investment of funds and time.

## 1. Introduction

Digital Single Lens Reflex (DSLR) cameras have been successfully used by astrophotographers since the advent of these imaging devices. Shortly after their introduction, several studies examined the suitability of DSLRs for photometry. Because they are not designed for photometry, these cameras present a unique set of challenges. For instance, DSLR sensors are manufactured with a Bayer array of red, green, and blue filters placed over individual sensor pixels. Unfortunately, the center wavelengths of these filters do not match the wavelengths of standard Johnson filters. Furthermore, features that improve image quality like built-in software noise reduction also distort the true photometric signature of stars of interest. Fortunately, camera manufacturers have given consumers access to the RAW pixel data, which is often free of any on-camera processing or noise reduction.

Examples of photometry testing of DSLRs include work by Hoot (2007). He tested a Canon EOS 350D as a stellar photometer by imaging a series of Landolt field stars. Using the Landolt V, B, and R filters, he computed offsets and color and extinction transformations. Hoot found significant ($\sim 0.13$ mag.) uncertainties in the color correction coefficients that were far in excess of the instrumental RMS ($\sim 0.003$ mag.) values. From this he concluded there must be some outlying systematic effect causing the low quality of fit. Furthermore he found that "no single exposure set taken with this DSLR family of camera can accurately span more than 2.5 magnitudes." Outside of this narrow window, the errors increase rapidly, therefore decreasing the utility of using DSLR cameras for fields with wide magnitude ranges.

Subsequent to publication of Hoot's article, DSLR cameras were put to the test on various astronomical targets. Littlefield (2010) and Guyon and Martinache (2011) have shown DSLR cameras are capable of 10 milli-magnitude or better photometry that is suitable for tracking transiting exoplanets. Fiacconi and Tinelli (2009) have shown DSLR cameras can track pulsating variable stars, like XX Cyg. All three of these studies made use of differential photometry that does not require a precise transformation to a standard photometric system. A recent study by Pata *et al.* (2010) shows that such transformations are indeed possible, provided that the spectral properties of the observed stars are known.

As part of the American Association of Variable Star Observers' (AAVSO) Citizen Sky Project, several participants elected to use DSLR cameras to track the 2009–2011 eclipse of ε Aurigae (Stencel 2008; Guinan and Dewarf 2002). The success of this method (Kloppenborg and Pearson 2011; Kloppenborg *et al.* 2011) inspired our work which confirms that DSLR cameras can indeed be used as accurate, photometrically calibrated, wide-field photometers over a wide range of colors and nearly five magnitudes of brightness with a very modest investment of time and equipment. What they lack in flexibility, DSLR cameras make up for in price and portability.

## 2. Procedure

2.1. Instrumentation and experimental setup

For our work we used a standard Canon 450D, without modifications to the filters, and two Nikkor lenses adapted to the Canon body. The 450D has a $22.2 \times 14.8$ mm, 12-megapixel CMOS sensor and is equipped with a 14-bit analog-to-digital converter. We verified the camera's linear response over most of its dynamic range by comparing sensor response to a series of illuminated targets. Target brightness levels were measured using a dedicated photometer. At ISO 100, the 450D appears to be linear (within ± 0.5 %) up to 14,300 ADU, where the camera sharply departs from the linear trend (full well saturation). The ADC clipping level occurred at ~15,800 ADU, rather than the expected 16,384 for a 14-bit camera. The calibration factor at ISO 100 has been measured to 2.27 e-/ADU. Therefore the 1 electron to 1 ADU calibration setting resides somewhere between ISO 200 and ISO 400. Above this setting, the dynamic range of the camera is reduced. We used standard Nikkor 200-mm and 85-mm lenses whose fields of view are $6.36 \times 4.24$ degrees and $15 \times 10$ degrees, respectively.

We measured the spectral response of our camera's Red (R), Green (G), and Blue (B) Bayer array filters. Figure 1a is a plot of the Canon 450D G filter compared to the Johnson's V filter definition (Maíz Apellániz 2006). We found the G channel is shifted blueward by 12 nm relative to Johnson's V-filter. A transformation equation is, therefore, required to adapt DSLR measurements to this photometric standard. Likewise, we found a 2-nm blue shift between Tycho VT and DSLR G (see Figure 1b), but this shift is nearly negligible. This result is quite interesting as it shows there should be little to no correction between the 450D G channel and Tycho VT. These properties will be verified and discussed in greater detail below.

Of note are some useful features of this camera to a photometrist. The camera has a $10\times$ magnified live-view display that is very useful when defocusing. Unfortunately, we found it is often not possible to frame the field of interest by using the live view at $1\times$ zoom. Instead we used the optical viewfinder with a right angle adapter added. Additionally, we used a red dot finder at times. To best simulate the facilities available to other observers, we mounted our camera on a small, undriven, equatorial mount to accelerate framing of the chosen fields. A simple tripod can also be used.

2.2. Choice of star fields

Our star fields were selected to optimize testing the limits of DSLR photometry and the related post-processing steps. In choosing our test field we intentionally excluded regions in the plane of the Milky Way to avoid blending target stars with potentially unseen, but detectable, background stars. Our four fields are found between the constellations Cygnus and Draco bounded by R.A. 19h 48m to 18h 05m and Dec. +48° 07' to +54° 28'. These fields total $15.85 \times 6.36$ degrees or 101 square degrees and have ~ 283 stars between V = 3.7 and 8.75 with (*BT*–*VT*) values of –0.2 to 2 (see Figures 2a and 2b). Of these, Vizier and VOPlot report 76 stars are at risk of blending within the photometric aperture and 27 stars are suspected variables (see Table 1). Most stars in the field are AFG and K spectral types, with a few B- and M-type stars (see Figure 2c). Because of the limited number of M-stars, we were not able to check the known transformation issues involving these objects (Perryman *et al.* 1997). We also obtained images from a second, larger region using the 85-mm lens. This area extended between R.A. $19^h 48^m$ to $17^h 10^m$ and Dec. $+44°$ to $+59°$ in three fields. These data have significant field distortions near the edge of the FOV that must be dealt with delicately. We will discuss these data and our method of reduction in a future publication. For the remainder of our discussion we shall refer to our observed fields as CD3S, CD4S, CD5S, and CD6S as summarized in Table 2.

2.3. Data acquisition

Our observations began on July 14, 2011, and ended on September 28, 2011. All imaging was done from the same location: $+47°$ 19' N, $+05°$ 01' E at 250 $m$ altitude. All data were taken in groups of 10 to 15 images (hereafter a "series") with the Canon 450D and Nikkor 200-mm lens at $f$/4, ISO 100 with 12.3-second exposure times, except a few of the CD3S field which uses only 8-second exposures. The number of images per series were chosen to reduce the noise from atmospheric scintillation to a few milli-magnitudes. All images were defocused slightly (see Figure 3) so that the stellar image covers roughly a 10- to 15-pixel diameter (plus trailing during the exposure). Most of the images were acquired within 15 to 30 degrees of the zenith where differential airmass is negligible. Four series for CD5S and CD6S were taken at 40 to 45 degrees of zenith at about 1.5 airmasses. From these images, we found that the brightest stars in our sample ($V \sim 3.7$) have a SNR $\gg 1,000$ and stars at 7th magnitude have SNR $\sim 200$.

Because we used a fixed focal length lens, it was possible to use a flat image recreated every few months. The flat is obtained by imaging a diffusely illuminated, non-glossy, fine-grained white surface (for example, the back of high quality photo paper). A 1% cross-surface uniformity of our flat fielding source has been verified using a dedicated photometer.

**3. Data reduction**

To reduce our data we employed two main stages. The first stage extracts the instrumental magnitudes and statistical information from the raw camera data, correcting for flat fielding, while the second stage calibrates the data to an absolute photometric system.

3.1. Stage 1

An automated reduction pipeline has been written in APL language running under a Dylog APL environment. This pipeline does the following:

1. Extracts the raw Bayer data in RGGB format from the camera's CR2 files using dcraw (from Dave Coffin, http://www.cybercom.net/~dcoffin/dcraw/).

2. The image is split into three planes formed by the R, $(G+G)/2$, and B pixels from the Bayer cell. This yields a $2145 \times 1428$ RGB image.

3. The Canon's systematic offset of 1024 ADU (possibly included in dark) is first subtracted from the raw. The resultant image is then flat fielded. Even though the image looks very uniform, we have noticed a small non-uniformity in the the outer perimeter of our images, and therefore we exclude a 50-pixel-wide border from photometric analysis. We do not apply a dark image as we have found these short exposures at ISO 100 have minimal dark current noise. Any residual cross-image gradients will easily be detected in stage 2, manifesting in the extinction coefficient.

4. Next we create a temporary luminance $(R+G+B)$ image to detect stars. The image is sampled at several points to determine the background and is fit using a polynomial. This function is then subtracted from the image, removing any remaining systematic background. The resulting luminance is again measured and pixels residing above 3-sigma (noise) above the dark level are selected as candidate objects. The area around the brightest pixels are analyzed to determine if they are part of a star. If confirmed, the footprint of the star is measured and its geometric centroid calculated to ensure proper centering during aperture photometry.

5. Because of diurnal motion, our star images are trailed. Therefore we perform aperture photometry using rectangles rather than annuli. Our inner aperture is $21 \times 13$ pixels and the outer is $51 \times 51$ pixels. The background level for each star, determined from the outer aperture, is subtracted from the foreground aperture and the RGB intensities of the star are extracted.

6. Next, stars are identified using the *Tycho 2* reference catalogue based upon image coordinates and a "tentative" instrumental magnitude relative to the ensemble of stars in the image is computed.

7. Lastly the star ID, RGB intensities (in electrons), R.A., Dec., and image position $(X, Y)$ are written to a file along with aggregate statistics like series mean altitude, means, standard deviations, and signal-to-noise ratio (SNR).

For our analysis we selected only stars whose SNR over a series is greater than 40. Below this limit, a star's SNR is no longer above the 3-sigma criterion for the automated extraction pipeline.

### 3.1.1. Stellar blending

One downside of wide-field DSLR photometry is that relatively low angular resolution of the camera can blend two nearby stars together. Upon examining our instrumental output table from the steps above, we found a significant number of outliers from the calibration trend we expected. Some of these stars were undoubtedly variables, but in most cases they were affected by blending with faint background stars that fell within the rectangular aperture.

To solve this issue we wrote a small piece of software that uses data from the *Tycho 2* catalog down to *VT* = 13. All stars that fall within the aperture are selected and their approximate flux is computed. If the background star contribution exceeds 0.012 magnitude, we flag it in our data tables. If the Point Spread Function (PSF) response of the camera were better characterized, we speculate it would be possible to remove the background star contribution, but this was beyond the scope of this paper. Any stars that were affected by blending were not used in our analysis.

### 3.1.2. Variable stars

In a similar manner to blended stars, we have also flagged variable stars. We have collected aggregate statistics on variability, spectral types, and luminosity from the *Tycho 2* (Høg *et al.* 2000a, 2000b), *Hipparcos Input* (Turon *et al.* 1993), *Hipparcos Main* (Perryman *et al.* 1997), and the *Tycho 2 Spectral Types* (Wright 2003) catalogues. Of our targets, twenty-four were flagged as variable stars from the input catalogs; however, six of these stars showed no sign of variation within our measured accuracy.

### 3.1.3. Final star selection and output quality

These observations have provided 201 stars from the four fields from VT magnitude 3.8 to 8.8. Of these, 67 have been deselected (18 variables, 52 blended), yielding a total of 134 stars for further processing. Aggregate statistics of these stars are shown in Figures 2a, 2b, and 2c.

### 3.2. Stage 2

The second stage of data reduction mirrors the techniques employed in the Citizen Sky Intermediate Reduction Spreadsheet. This method transforms the DSLR instrumental G magnitude (denoted using $\nu$ hereafter) into the standard photometric system (denoted using capital letters *V* and *B*) using the standard transformation coefficient method (compare to Henden and Kaitchuck (1982) and references therein). This method essentially fits the observed instrumental magnitudes to a 3D surface to determine the transformation coefficient ($\varepsilon$), extinction coefficient ($k'$), and zero-point offset ($\zeta_\nu$). The remainder of this section reviews this method by outlining the mathematics required to find these parameters.

### 3.2.1. Determining $\varepsilon$ and $\zeta_\nu$

For differential photometry in which airmass may be neglected, the transformation coefficient ($\varepsilon$) and zero point offset ($\zeta_\nu$) may be determined using the following equation (Henden and Kaitchuck 1982):

$$(V - \nu)_i = \varepsilon (B - V)_i + \zeta_\nu, \tag{1}$$

where *B* and *V* are the catalog B-band and V-band magnitudes, $\nu$ is the observed instrumental magnitude, $\varepsilon$ is the transformation coefficient, and $\zeta_\nu$ is the zero-point offset of the camera. The subscript $i$ denotes the $i^{th}$ calibration star in the image. Because of the way in which CMOS sensors are manufactured, we assume, to first order, that the response of each pixel in the camera is nearly identical. Therefore, if proper background and flat subtraction methods have been applied $\varepsilon$ and $\zeta_v$ are constant across the field. We may then either solve this equation graphically or use a linear least-squares fit (see Paxson 2010) in order to determine $\varepsilon$ and $\zeta_v$.

After the coefficients are determined, the above equation may be rearranged and the V-band magnitude for the $j^{th}$ target star computed via:

$$V_j = v_j + \varepsilon (B-V)_j + \zeta_v, \tag{2}$$

3.2.2. Airmass corrections

The above method of calibrating is good for images of small angular extent (that is, those with < 3˚ FOVs) at zenith angles less than 34 degrees. Beyond this point, the differential airmass across the field can contribute significantly to the error. First order airmass corrections may be applied to DSLR images using the following equation (Henden and Kaitchuck 1982):

$$(V-v)_i = -k'_v X_i + \varepsilon (B-V)_i + \zeta_v \tag{3}$$

where the newly introduced variable, $k'_v$, is the extinction coefficient and $X_i$ is the airmass. This equation has the same functional form as a geometric plane in three dimensions: $z = Ax + By + C$. If we assume that the instrumental magnitude, $v_i$, depends only on the terms on the right side of the above equation, then we may solve the above expression for the coefficients ($-k'_v$, $\varepsilon$, and $\zeta_v$) using a minimum of three calibration stars in the field of view. However, if one calibration star is incorrectly identified or the airmass is incorrectly computed, the coefficients will be skewed and the resulting magnitudes for target stars will be invalid. Therefore we alleviate this problem by using multiple calibration stars to compute the coefficients.

A least-squares fit of $n$ calibration stars to the plane defined by the equation $z = Ax + By + C$ is found by solving for the coefficient matrix, **X**, in following expressions, using the inverse of **A**:

$$\mathbf{AX = B} \tag{4}$$

$$\begin{bmatrix} \sum_{i=1}^{n} x_i^2 & \sum_{i=1}^{n} x_i y_i & \sum_{i=1}^{n} x_i \\ \sum_{i=1}^{n} x_i y_i & \sum_{i=1}^{n} y_i^2 & \sum_{i=1}^{n} y_i \\ \sum_{i=1}^{n} x_i & \sum_{i=1}^{n} y_i & \sum_{i=1}^{n} 1 \end{bmatrix} \begin{bmatrix} -k'_v \\ \varepsilon \\ \zeta_v \end{bmatrix} = \begin{bmatrix} \sum_{i=1}^{n} x_i z_i \\ \sum_{i=1}^{n} y_i z_i \\ \sum_{i=1}^{n} z_i \end{bmatrix} \tag{5}$$

It is not necessary to write a computer code to solve these equations, as many spreadsheet programs and programming languages already have built-in routines for such a purpose. For example, EXCEL/OPENOFFICE CALC have the "linest" function which we have employed in the Intermediate Reduction Spreadsheet on the Citizen Sky website. If you wish to write your own reduction code, Python's "scipy.optimize.leastsq" function can be used for this task.

After the coefficients are determined, the magnitude of the $j^{th}$ star in the field of view may be determined by rearranging Equation 6:

$$V_j = v_j + -k'_v X_j + \varepsilon (B-V)_j + \zeta_v \tag{6}$$

Note that these equations require that the color of the target stars must be known *a priori*! This places a DSLR camera at a significant disadvantage because even though the Blue and Red channels are measured, they do not correspond to any standard photometric filters. Furthermore, the spectral response of the Red and Blue pixels often are asymmetric with modest red and blue leaks compared to standard filters. This disadvantage can be mitigated by using a catalog that closely responds to DSLR G, thereby resulting in a near-zero value for $\varepsilon$ and mitigating the color contribution to final V-band output.

3.3. Verification

In addition to the APL-based pipeline described above, one of us (H.B.E.) created an alternative reduction method that uses AIP4WIN (Berry and Burnell 2005) to stack images, SOURCEEXTRACTOR (Bertin and Arnouts 1996) to automatically find and perform aperture photometry on stars, and SCAMP (Bertin 2006) to perform astrometric star association. The output is then processed in the same manner as step 2 described above using a script we have written in R (Matloff 2011). This second pipeline produced results identical (within uncertainties) to the method described above. Please contact H.B.E. if you are interested in virtual machine image of the freely redistributable components of this pipeline.

**4. Results**

Our data set consists of nearly 500 images taken in groups of ten, 12-second exposures. These 40 series represent nearly 80 minutes of combined exposure time. From the input catalog of approximately 300 stars brighter than $V = 8.8$, we detected about 200 stars in our fields that are 3-$\sigma$ above background noise. From these, we selected 134 stars that were free from blending and not identified as variables (or suspected variables) in catalogues. Overall, stars with $3.5 < V < 7.5$ had a mean uncertainty $< 0.01$ magnitude or better. Stars in the range $7.5 < V < 8.0$ had an average uncertainty of $\sim 0.02$ magnitude. Beyond this magnitude, photometric uncertainties rapidly grew $\propto 1/SNR$. We show typical results for constant and variable stars in Figure 4.

4.1. Choice of calibration catalogue

Following our discussion at the end of section 3.2.2 we have explored the use of one homogeneous photometric catalog and one inhomogeneous source for stage 2 calibration. In the next few paragraphs we describe the benefits of using a standardized system and caution against using inhomogeneous catalogs for calibration.

### 4.1.1. Tycho

For our first calibration catalog we used the *Tycho 2 Catalogue* (Høg *et al.* 2000a, 2000b). As discussed above, the difference in transmission between VT and DSLR-G filters is minimal. In Figure 5a and Figure 5b we adjusted DSLR-G to VT using only a zero point offset, $\zeta_v$. The residuals (that is, transformed–catalog) appear normally distributed about zero and show no systematic trends as a function of color, confirming the suspicion that the difference in transmission between the VT and RGB-G filters can be regarded as minimal for the Canon 450D.

### 4.1.2. ASCC

Unlike Tycho VT, the Johnson V measurements in the ASCC catalog (Kharchenko 2001) should exhibit a modest color transformation coefficient. For almost all of our target stars, ASCC contains VT magnitudes transformed to $V_J$. In Figure 6a we plot the instrumental magnitude $v_{inst}$ as a function of the ASCC2.5 Johnson V. To demonstrate the color correlation, the difference between instrumental and catalog magnitude is plotted as a function of $(B–V)$ in Figure 4b. Note that unlike our analysis for the VT photometry, and unlike the analysis in Hoot (2007), we find a significant correlation between the residuals and color.

Next we applied the full color calibration to the data to yield Figures 5c and 5d. Aside from a small excursion between $0.4 < (B–V) < 0.5$ (which contains only stars with $V > 8$ with very poor SNR), we find the data are normally distributed. Likewise, residuals as a function of magnitude appear normally distributed within an envelope that is $\propto 1/SNR$. Combined, these imply that the transformation equations for DSLR G to $V_J$ are valid across our entire range of colors and magnitudes in our sample.

### 4.1.3. SIMBAD

In Figures 5e, 5f, and 6c we essentially repeat the ASCC experiment using a heterogeneous reference catalog assembled from SIMBAD queries. Both the magnitude and color residual diagrams show a loss of precision of $\sim 0.005$ mag., with a few stars shifting by as much as 0.02 mag. Although certainly within the statistical uncertainties quoted in the catalog, these deviations are often outside of the internal instrumental uncertainty, implying the deviations are due to errors in the catalog magnitudes. We caution the reader that using aggregate catalogs, like blind queries from SIMBAD, may result in degraded precision. Therefore we suggest that DSLR photometry be performed using a standard reference catalog like ASCC or, preferably, Tycho VT.

## 5. Conclusion and discussion

We have shown that consumer-grade DSLR cameras can be used as accurate (0.01 mag.) photometers across a wide range of magnitudes and colors. In the next few paragraphs we provide a few comments concerning DSLR photometry and discuss how the reader may alleviate these issues.

5.1. On the number of reference stars

As discussed above, one may calibrate data using the simple method (Equation 2) using only two reference stars and the airmass-corrected version (Equation 6) with only three stars. However, when using these lower limits as a guide, the reader must be aware that an incorrect identification, bad instrumental magnitude, or incorrectly referenced catalog value will significantly degrade (if not invalidate) the results. As a rule of thumb, we recommend six to nine reference stars that bracket the target object(s) in color, airmass, and magnitude so that the values of $k'$, $\varepsilon$, and $\zeta_v$ may be interpolated rather than extrapolated.

In many cases, satisfying all of these requirements is not possible. The reader may be tempted to include fainter reference stars, but with that comes larger statistical uncertainty. In our work, we found stars with a SNR > 100 are often acceptable and strongly caution against using any star with a SNR < 100 as a calibrator.

5.2. Airmass corrections

Most of our images were taken fairly close to the zenith (with typical airmasses not exceeding 1.2), therefore the differential extinction across the image was negligible. Two series with the greatest airmass in field CD6S did show a significant correlation of residuals as a function of airmass (compare equation 6). Including airmass correction in the (planar) fit did improve the residuals significantly when compared to the (linear) color-corrected fit. In general, airmass corrections should be applied whenever the differential airmass across the entire FOV multiplied by the extinction coefficient exceed the desired level of accuracy.

5.3. Stellar blending

Of primary concern to a DSLR photometrist is the blending of foreground target/reference stars with fainter background stars. With the technique described above, stars can become blended in three different ways: (1) the formal resolution of the optical setup cannot resolve blended stars, even under ideal atmospheric conditions, (2) defocusing the image causes light from adjacent stars to overlap, or (3) diurnal motion causes an overlap of trails between two nearby stars. In all of these cases, if a comparison star is blended, it will skew the calibration and, potentially, invalidate the results. If the photometric target star(s) are blended, the measured photometry will certainly be invalid.

Before we consider these three cases in further detail, we wish to describe two methods by which blends might be identified. In a large ensemble of stars, blends will only have a limited effect on the resultant photometry. Blended stars can be detected in at least two ways. Given positions from an astrometric catalog and the formal resolution of the imaging setup, stars should be considered blended if the photometric extraction apertures overlap. Software that does this is available from author H.B.E. by request. If the reader wishes to use an empirical method for determining blends, blended stars may be identified by finding stars that skew the photometry error histogram. Typically a visual inspection of the plot that compares measured magnitudes to catalog data will have some obvious outliers. These are most frequently caused by blending, variability, or other sources of error (for example, small clouds).

The three sources of blending deserve further discussion. The method by which the photometrist chooses to decrease blending depends on the science objective they wish to achieve. In our work, a loss of 25% of our sample was inconsequential as we still obtained photometry on 150 stars with only ten 10-second exposures. Of the fifty-two stars that were lost due to blending, almost all of them were due to the limited resolution of the optics. We could have simply increased the angular resolution of our setup by zooming, but then we would require a larger number of exposures.

In the case where stars are blended due to defocusing or trailing, the method of resolving the blend becomes more difficult. In the case of defocus-induced blending, focusing the image is the obvious solution; however, defocusing ensures an accurate measure of the star's light. In the high photon regime, one should decrease the exposure length while proportionally increasing the number of exposures. When analyzing the data, the images should be aligned and stacked. Likewise, for bright stars that become blended due to trailing, the exposure length can be decreased while the number of exposures is proportionally increased.

This advice will likely not extend to the photon-limited, faint star regime. Indeed, the authors are unaware of any study that theoretically discusses the trade-offs between changing the intrinsic resolution, focus, and trail length while providing experimental verification of any published claims. Until such a work is completed, we suggest the reader carefully consider the science they wish to achieve and choose a setup best suited for the job. The procedure we describe herein is ideally suited for wide-field bright-star photometric monitoring, but clearly not for observing faint stars with little photometric variation.

## 6. Acknowledgements

The authors would like to thank the AAVSO's Citizen Sky project. B.K. acknowledges support from NSF grant DRL-0840188. B.K. thanks Dr. Doug Welch for discussions that piqued his interest in DSLR photometry.

The authors would also like to thank Arne Henden for answering questions concerning photometric calibration databases. We are grateful to the organizers of the Citizen Sky workshops I and II during which we first discussed these efforts. This research has made use of the SIMBAD database, operated at CDS, Strasbourg, France and several online collaborative tools like the AAVSO's IRC channel, Google Docs, and Pastebin.

*Table 1. The 27 variable stars detected in our survey.[1]*

| Tycho Identification | VT (mag.) | Series Mean (mag.) | Typical SD (mag.) | Series Min. (mag.) | Series Max. (mag.) | No. Sigmas | Tycho Classif. |
|---|---|---|---|---|---|---|---|
| 3528-2121-1 | 8.077 | 7.979 | 0.015 | 7.954 | 8.011 | 2 | DA |
| 3529-1447-1 | 7.648 | 7.633 | 0.010 | 7.600 | 7.660 | 3 | U |
| 3533-2577-1 | 5.219 | 5.195 | 0.003 | 5.182 | 5.209 | 5 | 1U |
| 3534-302-1 | 7.450 | 7.290 | 0.010 | 7.284 | 7.300 | 1 | 1D- |
| 3536-1939-1 | 7.359 | 7.379 | 0.006 | 7.356 | 7.420 | 7 | M |
| 3536-2022-1 | 8.371 | 8.400 | 0.023 | 8.335 | 8.436 | 3 | C |
| 3538-2150-1 | 8.436 | 8.415 | 0.015 | 8.351 | 8.483 | 5 | |
| 3539-137-1 | 7.759 | 7.783 | 0.008 | 7.757 | 7.810 | 3 | CU |
| 3539-1700-1 | 6.831 | 6.797 | 0.004 | 6.702 | 6.884 | 24 | 1U*0.51 |
| 3539-2623-1 | 8.313 | 8.389 | 0.015 | 8.345 | 8.461 | 5 | 3R |
| 3548-2346-1 | 7.227 | 7.234 | 0.005 | 7.225 | 7.246 | 2 | U*2.79 |
| 3550-579-1 | 8.339 | 8.310 | 0.029 | 8.276 | 8.367 | 2 | U*1.25 |
| 3551-1744-1 | 7.459 | 7.136 | 0.005 | 7.103 | 7.158 | 7 | P |
| 3552-1543-1 | 8.438 | 8.464 | 0.014 | 8.427 | 8.513 | 3 | |
| 3552-394-1 | 8.000 | 8.078 | 0.011 | 7.999 | 8.223 | 13 | P |
| 3553-999-1 | 8.260 | 8.266 | 0.012 | 8.249 | 8.282 | 1 | U |
| 3554-100-1 | 7.753 | 7.743 | 0.007 | 7.699 | 7.780 | 6 | 1U |
| 3554-1071-1 | 6.014 | 6.071 | 0.003 | 6.039 | 6.109 | 13 | 5U |
| 3555-686-1 | 7.559 | 7.551 | 0.008 | 7.525 | 7.582 | 4 | U |
| 3564-1126-1 | 8.121 | 8.129 | 0.010 | 8.103 | 8.173 | 4 | U |
| 3564-3159-1 | 6.231 | 6.215 | 0.003 | 6.151 | 6.256 | 21 | 5U |
| 3569-331-1 | 8.117 | 8.118 | 0.012 | 8.061 | 8.161 | 5 | 1U |
| 3908-1123-1 | 7.652 | 7.626 | 0.009 | 7.620 | 7.631 | 1 | U |
| 3918-1829-1 | 5.867 | 5.932 | 0.003 | 5.921 | 5.943 | 4 | 3U |
| 3920-1660-1 | 8.451 | 8.401 | 0.014 | 8.367 | 8.431 | 2 | D |
| 3920-1971-1 | 3.884 | 3.873 | 0.002 | 3.861 | 3.880 | 6 | C5 |
| 3934-27-1 | 7.405 | 7.413 | 0.007 | 7.374 | 7.436 | 6 | U |

[1] *Most stars have min./max. values that are 2+ times the typical nightly standard deviation. Of particular interest are TYC 3536-2022-1 which was labeled as "stable" in the Tycho 2 input and main catalogs. Stars TYC 3538-2150-1 and TYC 3552-1543-1 have no variability designation. All three stars have no variability designation in SIMBAD. Tycho classifications are: S=Standard, C=Stable in input/main catalog, U, P, M, R = confirmed variables. Numbers in Tycho classifications indicate variation type, see Perryman* et al. *(1997), and Wright* et al. *(2003) for designations.*

Table 2. Summary of the observed fields and exposure information.

| Field* | Center | | Near TYC | No. | No. | No. | No. | Air-mass |
|--------|--------|--------|----------|-----|-----|-----|------|----------|
| | R.A. | Dec. | | Series | Images | Days | Stars | |
| | h  m | °   ' | | | | | V < 8.8 | |
| CD3S "A" | 19 35 | +51 16 | 3568-2325-1 | 21 | 245 | 9 | 56 | 1.0–1.14 |
| CD4S "B" | 19 09 | +51 23 | 3554-275-1 | 6 | 76 | 3 | 40 | 1.02–1.2 |
| CD5S "C" | 18 43 | +51 30 | 3539-1697-1 | 6 | 76 | 3 | 48 | 1.05–1.34 |
| CD6S "D" | 18 15 | +51 30 | 3537-1538-1 | 7 | 85 | 4 | 57 | 1.13–1.53 |

*Fields are 15.9 × 6.4 degrees. The instrumental magnitudes for each of the 40 series of observations are available from R.P. by request.*

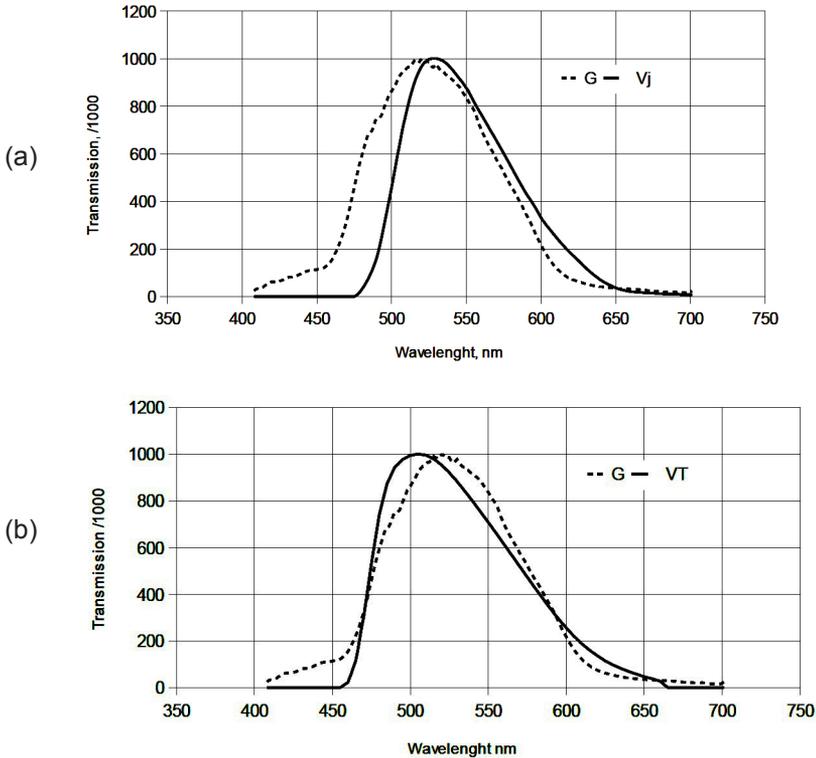

Figure 1. Instrumental response curves of the photometric channels (Photon-count) of the Canon 450D used in this experiment: (a) Johnson V and DSLR G; (b) Tycho VT and DSLR G.

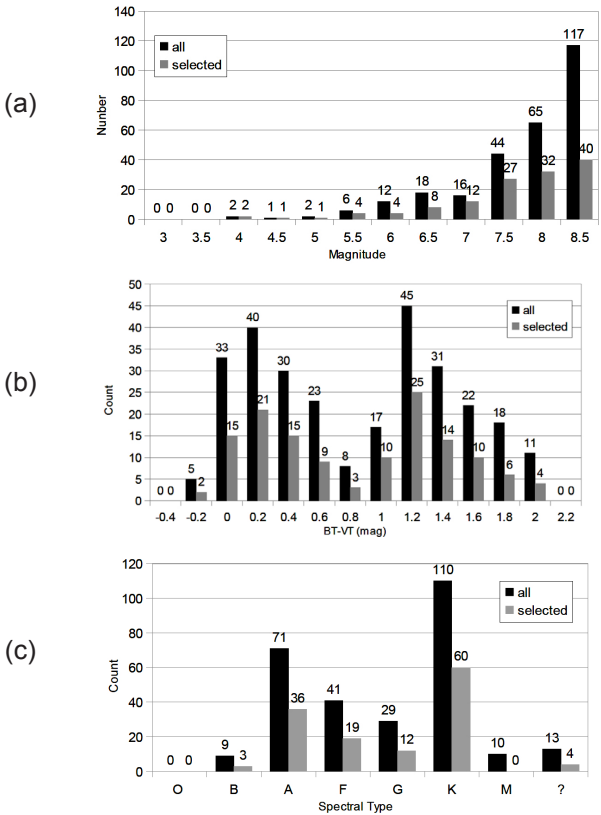

Figure 2. General field properties of the Cygnus-Draco 15.85 × 6.36 deg. fields, used in our experiment, and statistics for selected stars (after rejection of blended and variable stars): (a) magnitude ranges, Tycho 2; (b) (*BT–VT*) color ranges; (c) Approximate spectral types.

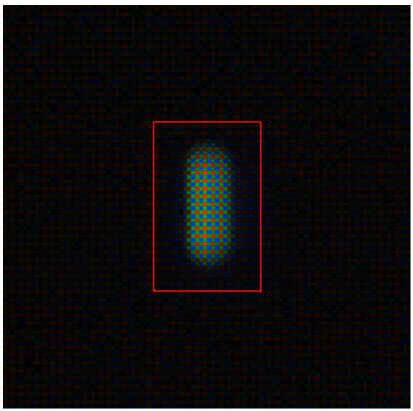

Figure 3. A typical RAW star image from our data set. As mentioned above, we use a non-tracking camera mount, hence the star image shows significant trailing. The Red Green Blue nature of the camera's Bayer array is clarly visible. The 21 × 13 pixel aperture is shown for reference.

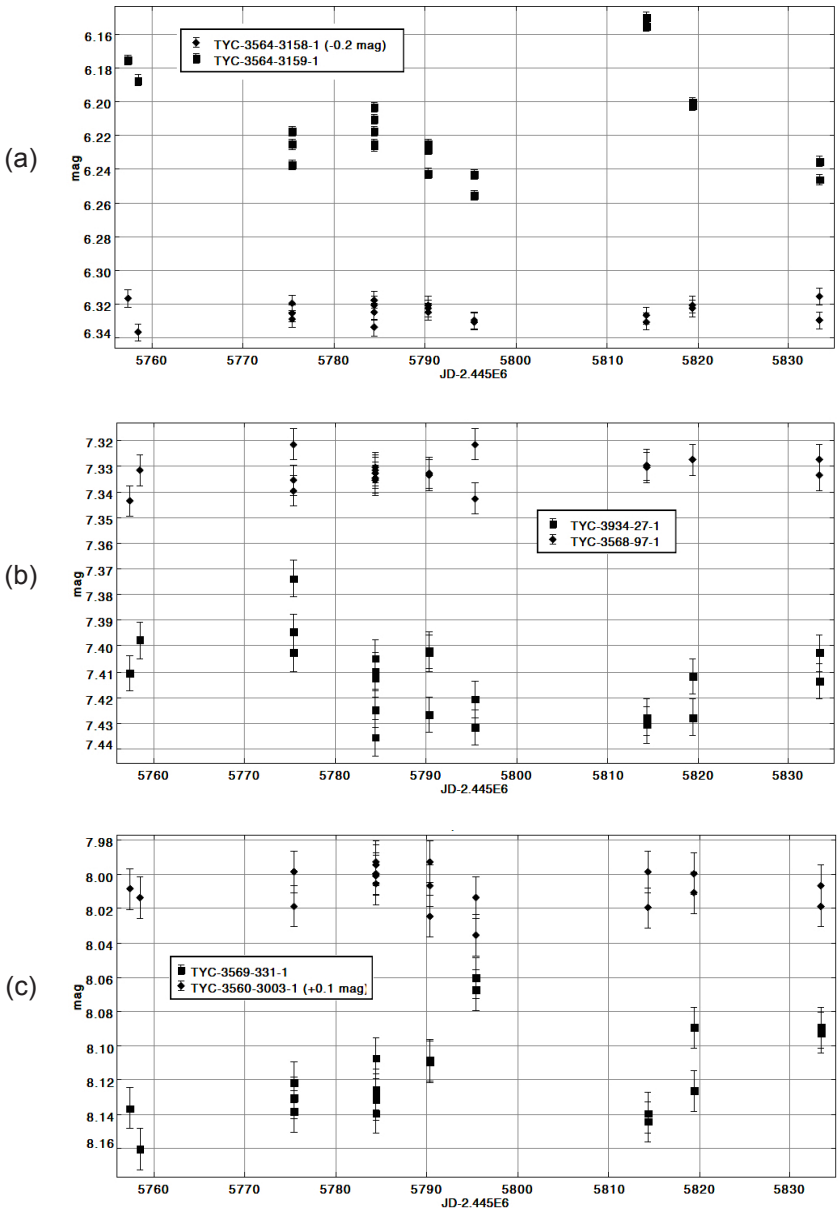

Figure 4. Examples of variable and non-variable stars seen in our data set: (a) nearly constant star TYC 3564-3158-1 and suspected 0.1 mag. variable TYC 3564-3159-1. (b) TYC 3568-97-1 and suspected 0.04 mag. variable TYC 3934-27-1. (c) 0.1 mag. suspected variable TYC 3569-331-1 and nearby stable star TYC 3560-3003-1.

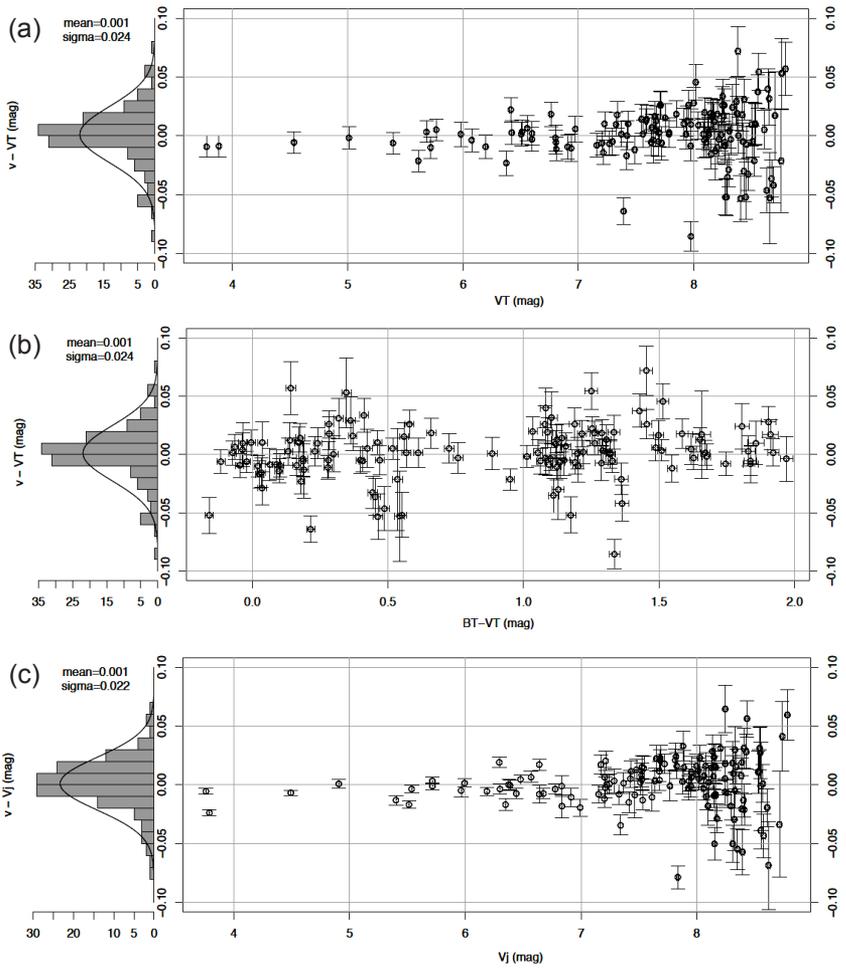

Figure 5 (a, b, c). Residuals of the simple calibration method as a function of color, magnitude, and catalog: (a) Tycho magnitude residuals; (b) Tycho color residuals; (c) ASCC magnitude residuals (figure continued on next page).

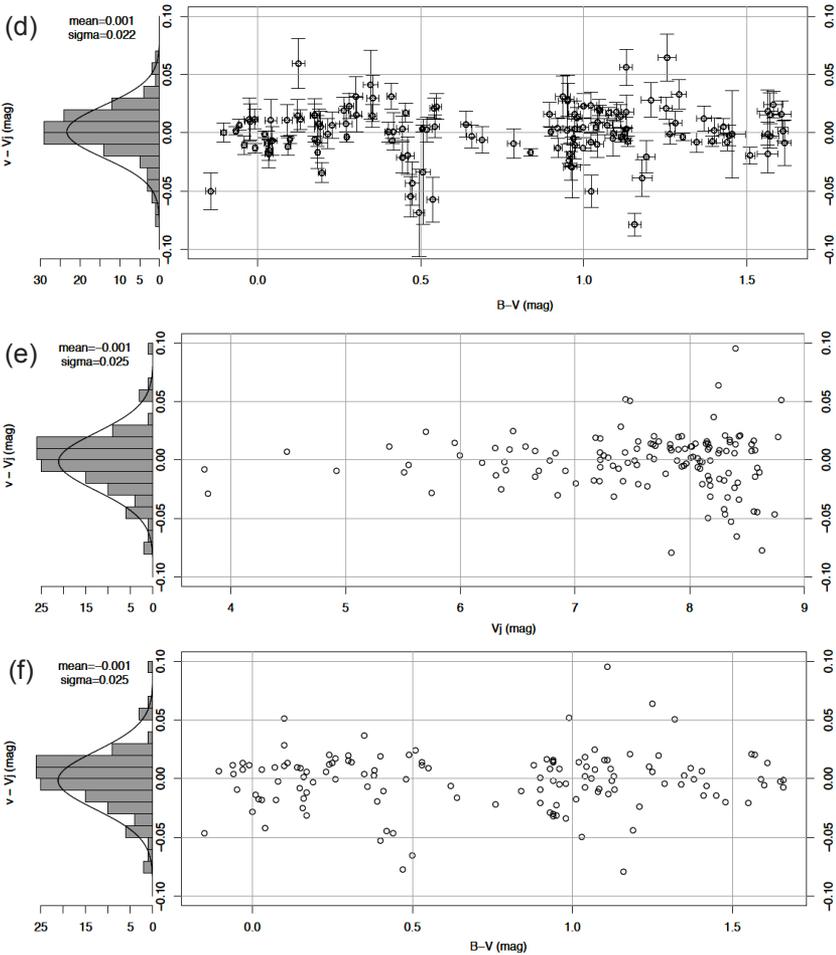

Figure 5 (d, e, f). Residuals of the simple calibration method as a function of color, magnitude, and catalog: (d) ASCC color residuals; (e) SIMBAD magnitude residuals; (f) SIMBAD color residuals. Aside from an increased spread at fainter magnitudes (due to a lower SNR), there appears to be no remaining systematic residuals after application of the calibration procedure discussed above. The statistical distributions of residuals are shown to the left of the $y$ axes. All residuals follow a Gaussian distribution with a very small offset ($\sim$ 1 mmag). The distribution is typified by a $\sigma = 22$ mmag uncertainty which improves by nearly a factor of two for stars with $V < 8$. This attests to the quality of the catalog and capabilities of the DSLR camera. We believe the SIMBAD results are skewed due to invalid color determinations inherent in a inhomogeneous catalog system.

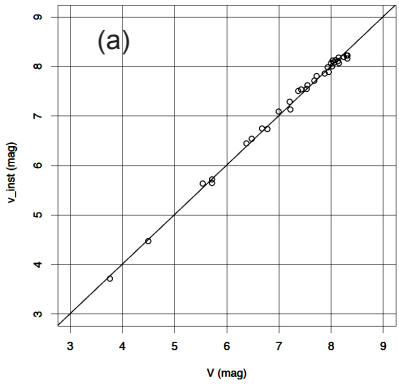

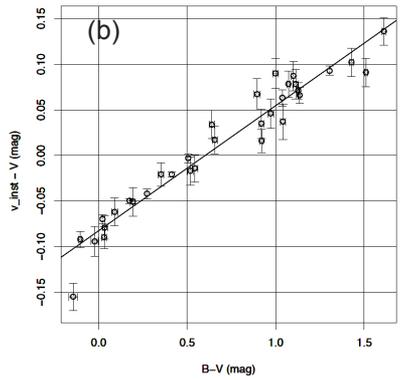

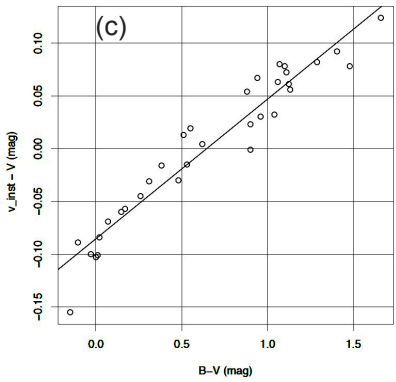

Figure 6. Raw instrumental magnitudes as a function of color and catalog magnitudes. These data show very linear trends, indicating the transformation methods described above are applicable: (a) ASCC2.5 instrumental magnitude vs. catalogue magnitude (single example series); (b) ASCC2.5 residuals vs. catalogue color, without color transformation (single example series); (c) SIMBAD residuals vs. catalogue color, without color transformation (single example series).